# The Spatial Sensitivity Function of a Light Sensor

N. K. Malakar[1], A. J. Mesiti[1], and K. H. Knuth[1, 2]

[1]Department of Physics, University at Albany, Albany, NY, USA
[2]Department of Informatics, University at Albany, Albany, NY, USA

**Abstract.** The Spatial Sensitivity Function (SSF) is used to quantify a detector's sensitivity to a spatially-distributed input signal. By weighting the incoming signal with the SSF and integrating, the overall scalar response of the detector can be estimated. This project focuses on estimating the SSF of a light intensity sensor consisting of a photodiode. This light sensor has been used previously in the Knuth Cyberphysics Laboratory on a robotic arm that performs its own experiments to locate a white circle in a dark field (Knuth et al., 2007). To use the light sensor to learn about its surroundings, the robot's inference software must be able to model and predict the light sensor's response to a hypothesized stimulus. Previous models of the light sensor treated it as a point sensor and ignored its spatial characteristics. Here we propose a parametric approach where the SSF is described by a mixture of Gaussians (MOG). By performing controlled calibration experiments with known stimulus inputs, we used nested sampling to estimate the SSF of the light sensor using an MOG model with the number of Gaussians ranging from one to five. By comparing the evidence computed for each MOG model, we found that one Gaussian is sufficient to describe the SSF to the accuracy we require. Future work will involve incorporating this more accurate SSF into the Bayesian machine learning software for the robotic system and studying how this detailed information about the properties of the light sensor will improve robot's ability to learn.

## INTRODUCTION

In this paper we study the spatial sensitivity function (SSF) of a light sensor used by a robotic arm developed in the Knuth Cyberphysics Laboratory to demonstrate Bayesian adaptive exploration [1]. The robotic arm deploys the light sensor to identify the center coordinates and radius of a white circle placed arbitrarily on a dark field. Previous implementations involved treating the light sensor as a point sensor, sensitive only to the light reflected from a single point below the detector. The sensor's response to a black or white stimulus was analyzed by comparison to a reference level, which was hard-coded into the inference algorithm. This posed obvious difficulties in the face of changes in the ambient light or in cases where measurements were taken near the edge of a circle. Such a crude implementation of the sensor response limits the ability of the likelihood function to adequately describe the data, especially when detecting object edges. Every light detector integrates the light arriving from a spatially distributed region within its field of view. The sensitivity of the sensor to light sources within this region is described by the SSF, which we will model using a parametric model defined by a mixture of Gaussians.

By characterizing the spatial sensitivity of the light sensor we aim to improve the quality of the robot's inference and inquiry processes.

## EXPERIMENT

The robotic arm uses the LEGO NXT Light Sensor, which consists of a photodiode placed alongside a red Light Emitting Diode (LED), which serves as a light source. Light emitted by the illuminating LED is reflected by the surface placed underneath it and the reflected intensity is recorded by the photodiode. The light sensor is connected to the LEGO Brick, which contains a 32-bit ARM7 ATMEL microcontroller. The Brick's internal software converts the photodiode's response to an integer between 1 and 100. We will refer to these as *LEGO units*, which for our purposes are the natural units to use since these are precisely the numbers that the robot obtains from the light sensor.

To obtain calibration data, we performed several measurements with known black and white patterns. These experiments were performed in a dark room to avoid complications due to ambient light and to eliminate the possibility of shadows cast by the sensor itself. The recorded SSF is not only a function of the properties of the photodiode, but also the illuminating LED.

Most experiments were performed using a surface background that is colored black on one half and white on the other. Figure 1c illustrates this pattern as a black-and-white strip; in the actual experiment it was much wider than the possible range of the sensor. The sensor was positioned 14 mm above the surface and moved horizontally from the black region to the white region (in the direction of the arrow in Fig. 1c) with intensities being recorded at millimeter intervals. Four sets of data were taken with the sensor in one of the four possible orientations at 90 degree intervals. Figure 1b shows the relative orientation of the sensor with corresponding symbols for the observed data points. Figure 2 shows the observed intensity as the sensor is moved from the black region to the white region. Note that the minimum intensity observation is not zero as the reflectance of the black surface is not completely zero. The asymmetric nature of the SSF is illustrated by the fact that the observed intensities do not follow the same curve at each orientation. One has to note that there are two frames of reference considered in this work: the sensor frame, which is defined with respect to the center of the sensor, and the lab frame, which is defined with respect to an arbitrary origin in the lab. The sensor frame coordinates $(x', y')$ are used to describe the SSF with respect to the sensor itself, while the response of the sensor to a black-and-white pattern will be depicted as a function of its position in the lab frame $(x, y)$. A proper transformation guarantees the correspondence between these two frames of reference.

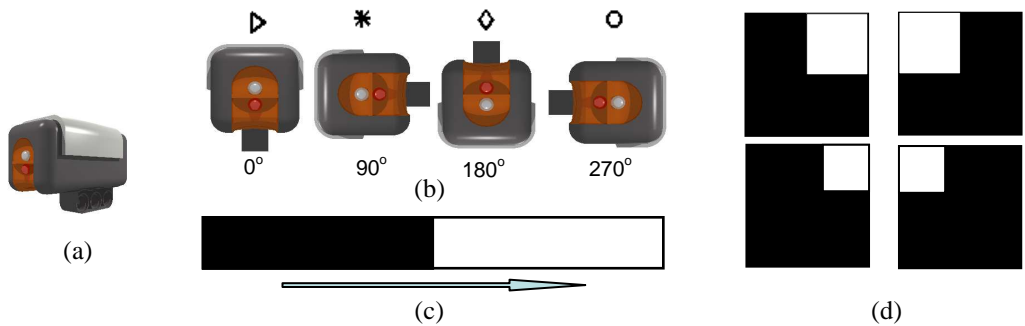

**FIGURE 1.** (a) The LEGO light sensor. The white circle is the photodiode, and the red circle is the illuminating LED. Note that they are separated by a narrow plastic ridge, which prevents the LED from shining directly into the photodiode. This ridge, along with the plastic lenses, and the presence of the illuminating LED will affect the SSF of the sensor. (b) The relative orientation of the sensor is shown with respect to the boundary separating the black and white observation field. Note the symbols corresponding to each orientation and the arrow showing direction of the motion as we take the measurements. These are used in subsequent figures displaying the data. (c) and (d) shows the patterns used in the calibration experiments.

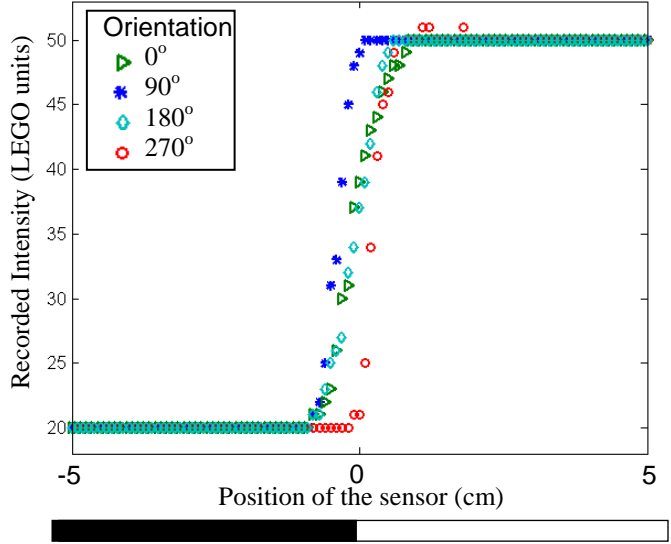

**FIGURE 2.** Observed intensity measurements for four orientations of the sensor showing the variation in the observed intensity as the sensor is moved across a dark field to bright field with a straight boundary as depicted in Fig. 1c. Figure 1b shows the orientations of the sensor corresponding to the symbols used above.

The black-white boundary alone does not give us sufficient data to infer a unique SSF since it has symmetries, which enable the orientation of the SSF to flip, or to change scale, and still produce identical predictions. To deal with this, we took four additional measurements using the patterns in Fig. 1d, which were designed to destroy the symmetries of the black-white boundary. The sensor was placed directly above the center of each pattern. In two cases (upper patterns), the corner of the white square was directly below the center of the sensor. In the other two cases (lower patterns), the center of the white square was shifted in both dimensions by 4 mm.

## BAYESIAN INFERENCE

We use a mixture of Gaussians (MOG) as a parameterized model to describe the spatial sensitivity function (SSF) in the sensor frame coordinates $(x', y')$

$$SSF(x', y') = \frac{1}{K} \sum_k a_k Exp\left[-\frac{1}{2}\left\{A_k(x' - u'_k)^2 + B_k(y' - v'_k)^2 + 2C_k(x' - u'_k)(y' - v'_k)\right\}\right], \quad (1)$$

where $(u'_k, v'_k)$ denotes the center of the $k^{th}$ Gaussian with amplitude $a_k$ and covariance matrix elements given by $A_k$, $B_k$, and $C_k$. The parameter $K$ denotes a normalization constant, which insures that the SSF integrates to unity.

The sensor's response to a known black-and-white pattern at the position $(x_o, y_o)$ in the lab frame can be modeled by

$$M(x_o, y_o) = I_{min} + (I_{max} - I_{min}) R(x_o, y_o), \qquad (2)$$

where $I_{min}$ and $I_{max}$ are observed intensities for a completely black surface and a completely white surface, respectively, and $R$ is a scalar response function that depends on the SSF. The minimum intensity $I_{min}$ acts as an offset and $(I_{max} - I_{min})$ serves to scale the SSF-based response. These parameters are estimated from the sensor's response to both a completely black and a completely white surface.

The sensor response function is given by an integral of the SSF-weighted albedo

$$R(x_o, y_o) = \int dx\, dy\ SSF(x - x_o, y - y_o)\, S(x, y) \qquad (3)$$

where $S(x, y)$ is the surface albedo in the lab frame, and $x' = x - x_o$ and $y' = y - y_o$ transforms position from the lab frame to the sensor frame. The result is a scalar quantity bounded between zero and one. In practice, we approximate this integral as a sum over pixels

$$R(x_o, y_o) = \sum_{x,y} SSF(x - x_o, y - y_o)\, S(x, y), \qquad (4)$$

and select $K$ in (1) so that the sensor response function is unity for a white background.

For the surface pattern in Fig. 1c, we define the origin of the lab frame to be at the center of the black-white boundary

$$S(x, y) = \begin{cases} 1 & \text{for } x > 0 \\ 0 & \text{for } x < 0 \end{cases} \qquad (5)$$

Similar albedos are defined for the four additional patterns.

The MOG model results in a set of parameters to be estimated $\boldsymbol{\theta} = \{a_k, u'_k, v'_k, A_k, B_k, C_k\}$ where the $k$ subscripts indicate a set of parameters with one for each Gaussian. Our goal is to use Bayes' theorem to infer the values of the set of parameters given by $\boldsymbol{\theta}$ based on the intensities $\mathbf{D}$ recorded during the calibration experiments

$$P(\boldsymbol{\theta} \mid \mathbf{D}, I) = P(\boldsymbol{\theta} \mid I) \frac{P(\mathbf{D} \mid \boldsymbol{\theta}, I)}{P(\mathbf{D} \mid I)}, \qquad (6)$$

where the symbol $I$ generically represents any prior information we possess. We assign a uniform prior for the model parameters over a reasonable range of the parameter values. Since we are ignorant about the uncertainties of the recorded intensities, we adopt a Student-t distribution as the likelihood function.

TABLE 1.

| Model: MOG | Log Evidence | Number of Parameters |
|---|---|---|
| 1 Gaussian | -665.5 ± 0.3 | 6 |
| 2 Gaussians | -674.9 ± 0.3 | 12 |
| 3 Gaussians | -671.9 ± 0.4 | 18 |
| 4 Gaussians | -706.1 ± 0.4 | 24 |
| 5 Gaussians | -999.6 ± 0.4 | 30 |

Five sets of measurements were taken. Four sets involved moving the sensor across the black-and-white boundary with the sensor in one of four orientations: $\phi = \{0°, 90°, 180°, 270°\}$. These orientations correspond to experiments $j = 1, 2, 3, 4$ respectively where $N_j = 101$ measurements were taken. The last set measurements, $j = 5$, involved positioning the sensor in the $\phi_5 = 0°$ orientation and presenting four patterns so that $N_5 = 4$. Assuming a Gaussian likelihood for the resulting intensity, but noting that the expected squared deviation of the recorded signal from the mean for each separate experiment, $\sigma_j$, is unknown, we integrate the Gaussian likelihood over $\sigma_j$ to obtain a Student-t distribution

$$P_j(D_j | \theta, I) = \left[ \sum_{i=1}^{N_j} \left( M_{\phi_j}(x_{ji}, y_{ji}) - D_j(x_{ji}, y_{ji}) \right)^2 \right]^{-N_j/2} \quad (7)$$

where $(x_{ji}, y_{ji})$ is the position of the sensor in the lab frame during the $i^{th}$ measurement of the $j^{th}$ experiment, $D_j$ refers to the measurements taken during the $j^{th}$ experiment, and $M_{\phi_j}$ refers to the predicted measurement during the $j^{th}$ experiment keeping in mind that the SSF has been rotated by $\phi_j$ degrees. The overall likelihood is the product of likelihoods for all five experiments, which we write here as a log likelihood

$$\log P(D | \theta, I) = -\frac{1}{2} \sum_{j=1}^{5} N_j \log \left[ \sum_{i=1}^{N_j} \left( M_{\phi_j}(x_{ji}, y_{ji}) - D_j(x_{ji}, y_{ji}) \right)^2 \right]. \quad (8)$$

We employed nested sampling [5] to explore the posterior probability for a series of MOG models, where the number of Gaussians in the mixture varied from one to five. The nested sampling algorithm is designed to compute both the mean posterior probability as well as the evidence. The algorithm is initialized by randomly sampling SSF models from the prior. The algorithm contracts the distribution of samples around high likelihood regions by discarding the sample with the least likelihood, $L_{worst}$. To keep the number of samples constant, another sample is chosen at random and duplicated. This sample is then randomized by taking Markov chain Monte Carlo steps subject to a hard constraint so that its move is accepted only if the new likelihood is greater than the new threshold, $L > L_{worst}$. This ensures that the distribution of samples remains uniformly distributed and that new samples have likelihoods greater than the current likelihood threshold. This process is iterated until convergence. The logarithm of the evidence is given by the area of the sorted log likelihood as a function of prior mass. When the algorithm has converged one can compute the mean parameter values as well as the log evidence. We ran the nested sampling algorithm for each of the five MOG models. The algorithm was initialized with 300 samples and iterated until the changes in log evidence were negligible. The resulting log evidence and the average weighted SSF for each of the five models was computed.

## RESULTS

The log evidence found for each MOG model is tabulated in Table 1 above. The model composed of only one Gaussian resulted in the greatest evidence with the two and three Gaussians model following close behind. Figure 3 illustrates the resulting SSFs for four of the explored models. Note that the SSF models for the first three models are extremely similar. Other models such as SSF with symmetric Gaussians as components and Gaussians without correlation elements (not reported) typically resulted in less evidence, which suggested that they did not have sufficient flexibility to describe the SSF. Figure 4 shows the original data along with the intensity measurements predicted by the one Gaussian model.

## DISCUSSION

This project focused on modeling the SSF of a light sensor with a mixture of Gaussians model. Intensity measurements were recorded in a set of five calibration experiments and used to explore the posterior probability of a set of MOG models using a nested sampling algorithm. The results indicated that a single two-dimensional Gaussian was sufficient to describe the data. These data were recorded at a particular height above the surface and it is possible that a different model order may be found to be more effective to describe the sensor response at different heights.

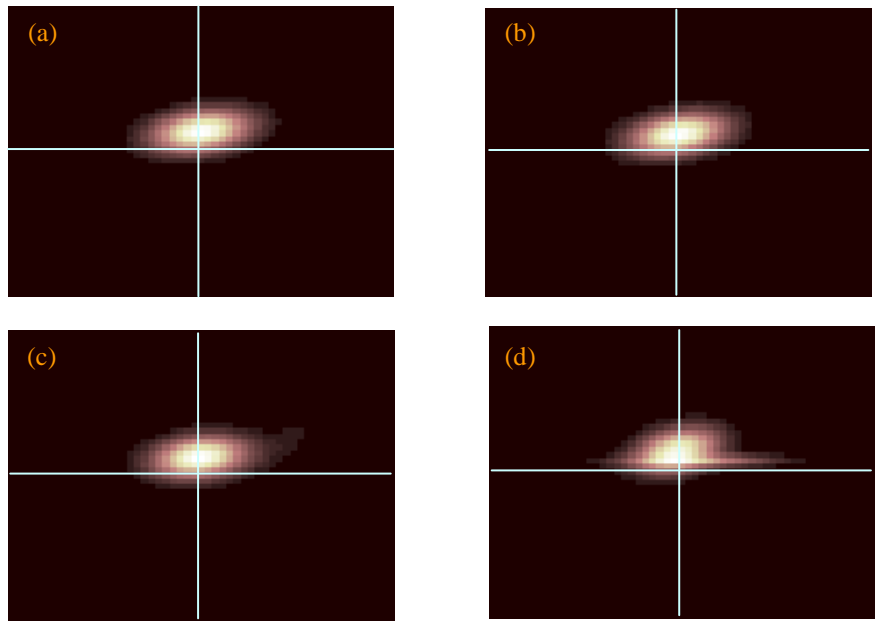

**FIGURE 3.** Average weighted SSF from MOG model with asymmetric Gaussians as the components: (a) One Gaussian (b) Two Gaussians (c) Three Gaussians (d) Four Gaussians. Axes represent the sensor frame.

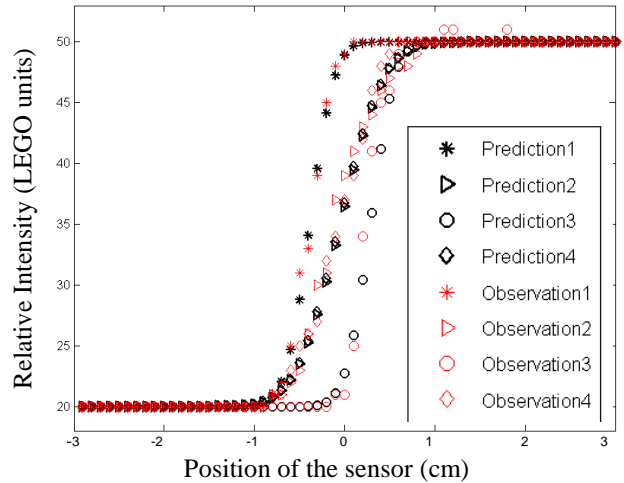

**FIGURE 4.** Comparison between predicted and observed intensities for four orientations of the sensor for the SSF model with a single Gaussian. The symbols correspond to orientations in Fig.1.

Future work will examine changes in the SSF as a function of height above the reflecting surface. In addition, rather than using a straight black-white boundary, we will instead consider a known patterns of white and black squares to get more information about the SSF from a given measurement. Given our current models of the SSF we could use Bayesian Adaptive Exploration [6] to select the most effective black-and-white surface patterns for calibration experiments.

## ACKNOWLEDGEMENTS

This research was supported in part by the University at Albany Faculty Research Awards Program (Knuth) and the University at Albany Benevolent Research Grant Award (Malakar). We are also thankful for the valuable comments and suggestions from the referees.## REFERENCES

1. Knuth K.H., Erner P.M., Frasso S. 2007, "Designing Intelligent Instruments", In: Knuth, A. Caticha, J.L. Center, A. Giffin, C.C. Rodriguez (eds.), Bayesian Inference and Maximum Entropy Methods in Science and Engineering, AIP Conference Proceedings 954, American Institute of Physics, pp 203-211.
2. Sivia D.S., Skilling J., "Data Analysis: A Bayesian Tutorial", 2nd Ed. Oxford University Press, Oxford, 2006.